\documentclass[runningheads]{llncs}

\usepackage[T1]{fontenc}
\usepackage{graphicx}
\usepackage{color, wrapfig}
\usepackage{tabularx}
\usepackage{multirow}
\usepackage{caption}
\usepackage{subcaption}
\usepackage{amsmath}
\usepackage{float}
\usepackage[table]{xcolor}
\usepackage{marvosym}

\newcommand{\rev}[1]{\textcolor{black}{#1}}

\begin{document}
\title{Reproducibility in Machine Learning-based Research: Overview, Barriers and Drivers}
\titlerunning{Reproducibility in Machine Learning-based Research}

\author{
Harald Semmelrock\inst{2} \and
Tony Ross-Hellauer\inst{1} \and 
Simone Kopeinik\inst{1} \and
Dieter Theiler\inst{1} \and
Armin Haberl\inst{3} \and 
Stefan Thalmann\inst{3} \and 
Dominik Kowald$^{\textrm{(\Letter)}}$\inst{1,2}}

\institute{Know Center Research GmbH, Graz, Austria \\\email{{tross,skopeinik,dtheiler,dkowald}@know-center.at} \and
Graz University of Technology, Graz, Austria \\\email{h.semmelrock@alumni.tugraz.at} \and
University of Graz, Graz, Austria \\\email{{armin.haberl,stefan.thalmann}@uni-graz.at}
}

\authorrunning{Semmelrock, Ross-Hellauer, Kopeinik, Theiler, Haberl, Thalmann, \& Kowald}

\maketitle

\begin{abstract}
\rev{Many research fields are currently reckoning with issues of poor levels of reproducibility. Some label it a ``crisis'', and research employing or building Machine Learning (ML) models is no exception. Issues including lack of transparency, data or code, poor adherence to standards, and the sensitivity of ML training conditions mean that many papers are not even reproducible in principle. Where they are, though, reproducibility experiments have found worryingly low degrees of similarity with original results. Despite previous appeals from ML researchers on this topic and various initiatives from conference reproducibility tracks to the ACM's new Emerging Interest Group on Reproducibility and Replicability, we contend that the general community continues to take this issue too lightly. Poor reproducibility threatens trust in and integrity of research results. Therefore, in this article, we lay out a new perspective on the key barriers and drivers (both procedural and technical) to increased reproducibility at various levels (methods, code, data, and experiments). We then map the drivers to the barriers to give concrete advice for strategies for researchers to mitigate reproducibility issues in their own work, to lay out key areas where further research is needed in specific areas, and to further ignite discussion on the threat presented by these urgent issues.}

\keywords{Machine Learning \and Artificial Intelligence \and Reproducibility \and Irreproducibility}
\end{abstract}

\section{Introduction}
\label{chap:introduction}
\rev{Trustworthy AI requires reproducibility~\cite{kowald2024establishing}. Unreliable results risk hindering scientific progress by wasting resources, reducing trust, slowing discovery, and undermining the foundation for future research~\cite{munafo_manifesto_2017,gundersen_state_2018}. However, many scientific fields currently face crucial questions over the reproducibility of research findings~\cite{baker_1500_2016}. Concerns of a ``reproducibility crisis'' have been most prominently raised in biomedical ~\cite{freedman2015economics,ioannidis2005most,iqbal2016reproducible,errington2021investigating} and social ~\cite{nosek2022replicability,open2015estimating,hardwicke2020empirical} sciences, but research employing Artificial Intelligence (AI) in general, and Machine Learning (ML) in particular, is also under scrutiny~\cite{hutson_artificial_2018}. ML is becoming ever more deeply integrated into research methods, not just in computer science but across disciplines~\cite{dwivedi2021artificial,ooi2023potential}. Indeed, recipients of the 2024 Nobel Prizes for both chemistry and physics included ML researchers. Hence, issues regarding the reproducibility of ML raise urgent concerns about the reliability and validity of findings not only for computer scientists but for large swathes of cutting-edge scientific research across disciplines.}

\rev{The causes of poor reproducibility can be technical, methodological, or cultural. Viewed from a high level, some causes, such as lack of sharing data and code, lack of or poor adherence to standards, suboptimal research design, or poor incentives, may be seen as common to many domains. Apart from the common challenges faced by other disciplines, the use of ML introduces unique obstacles for reproducibility, including sensitivity to ML training conditions, sources of randomness~\cite{raste_quantifying_2022}, inherent nondeterminism, costs (economic and environmental) of computational resources, and the increasing use of Automated-ML (AutoML) tools~\cite{koenigstorfer_black_2024,Haberl2025}. Among the methodological and cultural aspects, specificities of ML research, like  ``data leakage'', as well as ML-specific issues regarding unobserved bias, lack of transparency, selective reporting of findings, and publishing cultures, each play a role as well. Indeed, this cultural aspect must not be underestimated. The culture of ``publish or perish'' pervades academia, pushing researchers to publish as many papers in the highest-ranked or most prestigious journals or conferences as possible~\cite{pontika2022indicators}. In turn, this culture distorts incentives towards corner-cutting, giving rise to so-called ``questionable research practices'' and ``design, analytic, or reporting practices that have been questioned because of the potential for the practice to be employed with the purpose of presenting biased evidence in favor of an assertion''~\cite{banks2016evidence}.}

\rev{This paper aims to provide a detailed overview of reproducibility and its associated barriers and drivers in ML. This is urgently needed since, despite previous appeals from ML researchers on this topic, various initiatives from conference reproducibility tracks to the ACM's new Emerging Interest Group on Reproducibility and Replicability, and an expanding literature on the topic~\cite{mcdermott_reproducibility_2021,gundersen_machine_2022,heil_reproducibility_2021,kapoor_leakage_2022,gundersen_fundamental_2021,albertoni_reproducibility_2023,semmelrock2023reproducibility,gundersen2023sources}, we contend that the general community continues to take this issue too lightly. In addition, despite the growing literature, no such comprehensive overview exists.} For example, in~\cite{gundersen2023sources}, the authors identify and categorize sources of irreproducibility in ML and how these sources affect conclusions drawn from ML experiments. However, this study does not investigate the drivers to address these sources of irreproducibility. Thus, our paper provides a contextual categorization of the barriers and drivers to the four types of ML reproducibility (description, code, data, and experiment) proposed by ~\cite{gundersen_fundamental_2021}, with specific reference to research in both computer science and biomedical fields. We also propose a Drivers-Barriers-Matrix to summarize and visualize the results of the discussion. Such an analysis stands to clarify the current state regarding ML reproducibility, \rev{to give concrete advice for strategies for researchers to mitigate reproducibility issues in their own work, to lay out key areas where further research is needed in specific areas, and to further ignite discussion on the threat presented by these urgent issues.} 
The paper is structured as follows: in Section~\ref{s:definition}, we clarify terms and working definitions. We then analyze the barriers to increased reproducibility of ML-driven research  (Section~\ref{chap:barriers}), and next, the drivers that support ML reproducibility, including different tools, practices, and interventions (Section~\ref{chap:drivers}). \rev{Here, we also provide a comparison of the strengths and potential limitations of these drivers.} Finally, we map the barriers to the drivers to help determine the feasibility of various options for enhancing ML reproducibility (Section~\ref{chap:discussion}). We close the paper with a conclusion and an outlook into our future research in Section~\ref{chap:conclusions}. 

\section{Defining Reproducibility}
\label{s:definition}

The concept of reproducibility can have different interpretations across various research fields and even within the same field~\cite{fidler_reproducibility_2021}. 
To avoid confusion, we first specify our terms, broadly defining reproducibility and then further categorizing it into various types and degrees. The first distinction comes from Goodman et al.~\cite{goodman_what_2016}, who specify a fundamental division between whether we (i) mean reproducible in principle (termed ``methods'' reproducibility) due to sufficient description/sharing of methodologies, materials, etc., or (ii) whether results/conclusions actually prove to be reproducible when experiments or analyses are re-done. In the second category, they distinguish ``results'' and ``inferential'' reproducibility, depending on whether the analyses or inferences to broader conclusions are reproduced. 

Within ML research, widely accepted definitions that build further on these key distinctions have been proposed by Gundersen et al.~\cite{gundersen_machine_2022,gundersen_fundamental_2021}. We follow and build upon these latter definitions, and hence, here outline them at some length. 
Gundersen et al.~\cite{gundersen2023sources} define reproducibility in general as ``the ability of independent investigators to draw the same conclusions from an experiment by following the documentation shared by the original investigators''. 
Relating to point (ii) of Goodman et al.'s schema, Gundersen and colleagues ~\cite{gundersen_machine_2022,gundersen_fundamental_2021} further distinguish the targets of reproducibility, i.e., how closely an experiment can be reproduced:

\begin{itemize}
 \item \textbf{Outcome reproducibility} requires the reproduced experiment to have the same or adequately similar outcome as the original experiment. Due to this, the same analysis and interpretation follow, and the hypothesis is either supported or rejected by both experiments.
 \item \textbf{Analysis reproducibility} does not require the reproduced experiment to have the same/similar outcome; however, if the same/similar analysis and, therefore, also interpretation can be made, an experiment is analysis reproducible.
 \item \textbf{Interpretation reproducibility} does not require the reproduced experiment to have the same/similar outcome nor analysis but requires the interpretation to be the same as the original one.  
\end{itemize}

This categorization aims to overcome the problem of ambiguity when making specific claims about the reproducibility of an experiment. Often in literature, authors write about reproducing the same ``results'' of an experiment. It is not apparent, however, in which cases they mean to achieve the same computational outcome, i.e., outputs of the algorithms, or whether they mean to reach the same analysis or interpretation. Therefore, achieving interpretation reproducibility is a more general and often less stringent requirement than achieving outcome reproducibility. This categorization is not specific to ML, but is generally applicable to any research field that conducts data analysis and interpretation. 

In addition, relating to point (i) of the Goodman et al. schema (``methods'' reproducibility),~\cite{gundersen_fundamental_2021} specifies reproducibility types to which methods can be made transparent through description or sharing. These four types are defined as \textit{R1 Description}, \textit{R2 Code}, \textit{R3 Data}, and \textit{R4 Experiment}. The lower the level of reproducibility, the less shared information is shared, making the study more difficult to reproduce. 
As an example, in general, all published research experiments are accompanied by a textual description of the experiment. If this textual description is the only information shared by the authors, the research is categorized as \textit{R1 Description}, which, according to this scheme, is the minimal kind of reproducibility. In contrast, if all three building blocks, i.e., text, code, and data, are shared, the experiment can be categorized as \textit{R4 Experiment}, the most expansive kind of reproducibility. Furthermore, the distinction between \textit{R2 Code} and \textit{R3 Data} is defined by whether the textual description is accompanied by either code or data, respectively. 
\begin{figure}[!t]
	\begin{center}
		\includegraphics[width=.65\textwidth]{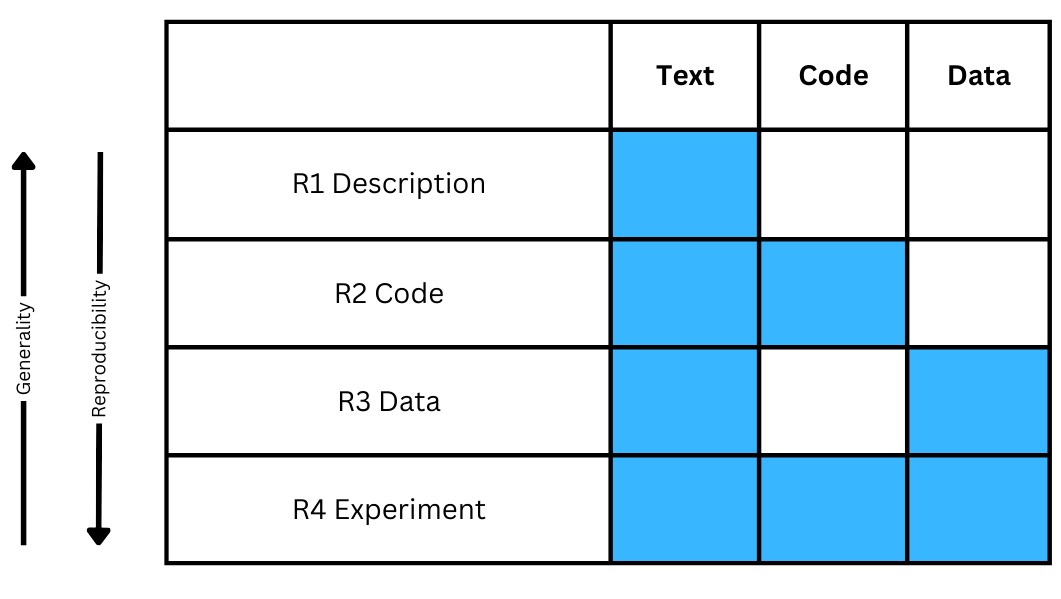}
	\end{center}
	\caption[Types of ML reproducibility]{\textbf{Types of reproducibility}. Adapted from Gundersen~\cite{gundersen_fundamental_2021}.\vspace{-6mm}}
	\label{fig:degrees}
\end{figure}
Also, the different types of ML reproducibility exhibit an interplay between generalizability and reproducibility. \textit{R1 Description} leads to strong generalizability but to weak reproducibility, while \textit{R4 Experiment} leads to stronger reproducibility but weaker generalizability. \rev{What this means in sum is that rerunning the same code on the same data using the same description will make it likelier to obtain the same results, but those results might still be wrong due to errors or biases in any of those elements. On the other hand, building code from scratch and using alternative datasets for analysis will show that the techniques give similar results across contexts and, hence, higher levels of confidence in the generalizability of findings.} These relationships are illustrated in Figure~\ref{fig:degrees}. In what follows, unless stated otherwise, we use the definitions proposed by Gundersen and colleagues~\cite{gundersen_fundamental_2021}. 

\section{Barriers in ML Reproducibility}
\label{chap:barriers}
Next, we discuss nine barriers to ML reproducibility, categorized into the four types of reproducibility mentioned beforehand. Where applicable, we give examples within the research fields of biomedical science and computer science. 

\subsection{R1 Description}
\subsubsection{Completeness and quality of reporting.}
Research often lacks reproducibility due to missing or vague methodological details. Mainly, there are three issues in this regard, which often hinder the reproduction of study results~\cite{pineau_improving_2021}:
\begin{enumerate}
    \item The ML model or training procedure is either incorrectly specified or under-specified. Reports should give clear details on all steps of the procedure, even if data and code are not shared. This includes details about which ML models are used, as well as details on the training data and data preprocessing. 
    \item The evaluation metrics used to report results are not properly specified. There are many metrics which can be used to evaluate ML models, e.g., accuracy, receiver-operator-curve (ROC), or mean-squared-error (MSE). It is important to define these metrics and also explain why they were used. 
    \item \rev{Often results are selectively reported, e.g., researchers may only provide results for the best test run  out of many test runs, instead of properly assessing and reporting average values and variances~\cite{belz_systematic_2021}.} 
\end{enumerate}
\noindent
Generally, it is important for studies to use a robust methodology and provide detailed reports so that other researchers can verify results and understand how analyses were conducted. While ML models have proven highly effective in biomedical fields, studies often fall short in providing comprehensive and high-quality reporting. For example, in studies on predicting cardiometabolic risk from dietary patterns~\cite{panaretos_comparison_2018} or supporting the clinical management of diabetes~\cite{kamel_rahimi_machine_2022}, ML models were observed to be very promising. While this further enhances the promise of applying ML models for various clinical prediction tasks, there is a clear need for thorough reporting and validation of these models to allow for their integration into routine clinical care~\cite{kamel_rahimi_machine_2022}. This is also true for the application of ML models for cancer imaging, where ML models often surpass radiologists in performance, but publications on these models lack the documentation details needed to reproduce the results~\cite{provenzano_radiologist_2021}.  

\subsubsection{Spin practices and publication bias.}
\label{section:spin}
Another issue commonly observed in ML-based research that negatively affects reproducibility is ``spin''. It refers to the misuse of language to ``intentionally or unintentionally affect the interpretation of study findings''. It is also understood as an inconsistency between the study results and the conclusions, in the sense that results are over-generalized or the claimed conclusions are not supported by the scientific method. This has been shown to impact both the interpretations and decision-making by readers~\cite{andaur_navarro_systematic_2023}. 
In ML-based biomedical research, the most common practice of spin includes recommending models for various applications without providing external validation in the same study. More concretely, the recommendation to use a model either in a clinical setting or for a different population is only validated in approximately 15\% of the cases. Other observed instances of spin are invalid comparisons of results to previous studies and the use of leading words and strong statements to make the results sound more significant~\cite{andaur_navarro_systematic_2023}.
\rev{The prevalence of spin can perhaps be attributed in large part to the academic culture of ``publish or perish'' and its associated reward systems. Valuing and rewarding perceived novelty and potential impact over basic rigor and responsible reporting can lead researchers to inflate claims in hopes of acceptance in the most prestigious venues. It can also skew the literature in other ways, leading to so-called ``publication bias''~\cite{saidi2024unravelingoveroptimismpublicationbias}. Here, in addition to spin and the aforementioned selective reporting of (usually positive) results, the role of the peer review system is also in question, given known biases on the part of reviewers that can lead to preferential treatment for researchers from specific regions, institutions or demographics, or for certain types of research~\cite{lee2013bias}. Other kinds of bias, such as ``complexity bias'' (tendency to prefer complicated over simple results and explanations), are also known to influence acceptance decisions~\cite{trosten2023questionable}.}

\rev{Finally, we note how a core aspect of computer science culture may exacerbate these issues, namely the importance of conference rankings. Within computer science, the prime mode of publication is within conference proceedings, with conferences ranked A* to C by bodies such as ICORE\footnote{\url{https://portal.core.edu.au/conf-ranks/}}. 
To date, surprisingly little has been said regarding the analogous nature of such conference rankings to other metrics like the Journal Impact Factor, where a rich literature exists critiquing its worth as an indicator of quality or impact for individual pieces of work~\cite{lariviere2019journal}.  
Although elaboration at length on this issue is outside the scope of this article, we suggest this as an underexplored topic for future research. Such research can build on a rich evidence-base exploring downstream ill-effects of badly designed or misused metrics, including distortion of incentives~\cite{smaldino2016natural}, inviting manipulation or gaming~\cite{biagioli2020gaming}, goal displacement and task reduction~\cite{rijcke2016evaluation}, and influencing core academic values~\cite{burrows2012living}.
}

\subsection{R2 Code}

\subsubsection{Limited access to code.}
Published ML research is often not accompanied by available data and code. Only one-third of researchers share data, and even fewer share their source code~\cite{hutson_missing_2018}. This can be attributed to several factors, such as the increasing pressure on researchers to publish quickly, which often leaves insufficient time to refine code and decreases the willingness to share it. Additionally, concerns about intellectual property may further discourage researchers from releasing their code.
According to Gundersen and Kjensmo~\cite{gundersen_state_2018}, sharing ML code to facilitate reproducibility requires publishing seven pieces of information: hypothesis, prediction method, source code, hardware specifications, software dependencies, experiment setup, and experiment source code. Unfortunately, current research rarely meets these requirements, leading to reproducibility issues due to different software versions, hyperparameter settings, or hardware differences~\cite{hong_evaluation_2013,belz_systematic_2021}. For example, research in recommender systems has struggled with the lack of shared code, significantly contributing to a lack of reproducibility. Even when code is shared, it is often incomplete, poorly documented, or limited to pseudocode or skeletal implementations rather than fully executable code~\cite{cremonesi_progress_2021}. To address this, shared code should encompass comprehensive documentation, including scripts for data preprocessing, hyperparameter tuning, management of random seeds, and implementations for comparisons against baseline models.

\subsection{R3 Data}

\subsubsection{Limited access to data.}
The main reproducibility barrier associated with \textit{R3 Data} is that data is simply not shared or made publicly available most of the time~\cite{hutson_missing_2018}). \rev{A review in biomedical research, specifically radiomics, investigated 257 recent ML publications and found that only 16 of them shared data or used publicly accessible datasets~\cite{d2024towards}}. This could be due to privacy concerns or a lack of incentives and motivation. Moreover, many benchmark and training datasets encounter challenges related to copyright, licensing, and longevity. These datasets may also raise ethical concerns, such as the unintentional inclusion of privacy-sensitive or harmful content, making it difficult to share the data for ML model training~\cite{paullada_data_2021}. 
Similar to the sharing of source code, sharing only the datasets is insufficient without sufficiently detailed levels of documentation. For proper use, it is also important to share specific splits, i.e., the training dataset, validation dataset, and test dataset~\cite{gundersen_state_2018}. Furthermore, data sharing needs to be accompanied by documentation specifying details about the provenance and preprocessing of data. \rev{Significant recent initiatives aiming at improvement here are: \textit{Croissant}, a unified format for machine learning datasets that integrates metadata, resource descriptions, data structure, and default ML semantics in a single file\footnote{\url{https://github.com/mlcommons/croissant}}, and \textit{MLCommons}, which is working towards open benchmarks and public data\footnote{\url{https://github.com/mlcommons}}. In addition, RO-Crate is a standard for packaging data and other research objects together with data to enable reuse and reproducibility\footnote{\url{https://www.researchobject.org/ro-crate/}}.} 

\rev{Standardising and mainstreaming these practices is essential for validation and checking of methods. We next discuss two common methodological errors related to data, data leakage and bias, and their impact on reproducibility.}

\subsubsection{Data leakage.} 
\label{subsection:data leakage}
In practice, methodological issues such as data leakage (also referred to as target leakage) often hinder the reproducibility of ML-based research~\cite{kapoor_leakage_2022}. This is due to the growing number of non-experts employing machine learning across different research fields~\cite{gibney_could_2022}, which is fueled by the ease of application of ML libraries and no-code off-the-shelf AI tools. In essence, data leakage happens when data on which the ML model should not be trained leaks into the training process. 
Data leakage can be categorized into 3 subcategories~\cite{kapoor_leakage_2022}: 
\begin{enumerate}
    \item No clean train/test split. Here, four variants are possible: (1) training data and test data are not split at all, (2) test data is also used to select the best features from the training data (feature selection), (3) test data is also used for imputation of missing data during preprocessing, and (4) duplicates occur in the training and test data. 
    \item Use of non-legitimate data. For example, when the use of antihypertensive drugs is used as a feature to predict hypertension. This data is non-legitimate, since it would not be available in a real-world scenario (people are prescribed those drugs because of a hypertension diagnosis) and would be useless for predicting hypertension in undiagnosed patients. 
    \item Test set is not drawn from the distribution of scientific interest. There are three possible variants: (i) temporal leakage, which is problematic for ML models that attempt to predict future outcomes, i.e., when some training samples have a later timestamp than samples available in the test set, (ii) the training and test data are not independent of each other, e.g., there should not be samples in the training and test data that are drawn from the same person, and (iii) the test set is not chosen selectively; for instance, if the model is solely evaluated on data, on which it performs well. 
\end{enumerate}

\subsubsection{Bias.}
\label{section:bias}
Bias in ML refers to the error introduced by approximating a real-world problem, which may be complex, by a simplified model, often leading to systematic deviations from the real world. Furthermore, biases can arise when the model contains imbalances or reflects existing societal biases~\cite{mehrabi2021survey}.  
ML models, which are subject to bias, are prone to generalization issues, and are therefore potentially problematic for ML reproducibility. 
There are eight kinds of bias that can arise during the data handling phase of ML development~\cite{rouzrokh_mitigating_2022}: (i) selection bias -  using data not being representative of the target group, (ii) exclusion bias - excluding particular data samples based on the belief that they are unimportant, (iii) measurement bias - favoring certain measurement results, (iv) recall bias - labeling similar data samples differently, (v) survey bias - introducing data issues stemming from data collection surveys, (vi) confirmation bias - favoring information, which confirms previous beliefs, (vii) prejudice bias - including human-related prejudices in training data, and (viii) algorithmic bias - replicating or amplifying biases by the inner workings of the algorithms.  
Bias, such as selection bias, often leads to the issue of validity shrinkage in biomedical science research~\cite{ivanescu_importance_2016}. For example, in obesity and nutritional research, ML is used to predict obesity, heart rate, or the risk of a heart attack based on data from an individual. 
Here, validity shrinkage refers to the issue that a predictive model trained on a subset of data will most probably not perform well on new samples. The difference between predictive performance on known data and new data, however, is most often not accounted for in nutritional science, and therefore also leads to performance claims that cannot be reproduced~\cite{ivanescu_importance_2016}. 

\subsection{R4 Experiment}

\subsubsection{Inherent nondeterminism.}
Inherent nondeterminism in ML models means that results can vary between test runs, even with identical code, data, and hyperparameters. This variation arises from sources of randomness during training, such as, e.g., random parameter initialization, stochastic optimization, and random data subsampling (e.g., in k-fold cross-validation) \rev{and the complex interactions between them, which are fundamental characteristics of most ML workflows~\cite{raste_quantifying_2022,leventi-peetz_deep_2022}}. 
Neural networks are especially known for their inherent nondeterminism, leading to varied computational outcomes in multiple reruns due to increased sources of randomness during training~\cite{ahmed_managing_2022}. 
In some cases, inherent nondeterminism can cause such large variations that reruns not only yield slightly different outcomes, but also lead to significant fluctuations in the performance of an ML model~\cite{ahmed_managing_2022} \rev{or to varying conclusions in ML model comparisons~\cite{gundersen2023reporting}}. 
This issue is exacerbated when other sources of variation, such as different hyperparameters, are introduced to the ML model. In such cases, the impact can be magnified, and it is often observed that minor changes in hyperparameters can result in significant performance loss~\cite{belz_systematic_2021}. Reviews of reproducibility in both NLP research~\cite{belz_systematic_2021} \rev{and biomedical research~\cite{ahmed_measuring_2022} highlight these core issues with nondeterminism that are exemplary for ML research because they reflect challenges that are prevalent across ML-based research and serve as representative examples of the broader reproducibility crisis faced by the ML community.}
Simply rerunning the original code of an experiment during a reproduction leads to large variances of results and different computational outcomes on each run.
Reinforcement learning, a subfield of ML, is particularly susceptible to these reproducibility issues, partially because of additional sources of nondeterminism, such as the reinforcement learning environment or policy~\cite{nagarajan_deterministic_2019}. 

\subsubsection{Environmental differences.}
Various studies have demonstrated that hardware differences, such as different GPUs or CPUs, and compiler settings can lead to different computational outcomes~\cite{hong_evaluation_2013}. Additionally, a comparison between the same ML algorithm with fixed random seeds executed using PyTorch\footnote{\url{https://pytorch.org/}} and TensorFlow\footnote{\url{https://www.tensorflow.org/}} resulted in different performances~\cite{pouchard_replicating_2020}. \rev{Furthermore, even different versions of the same framework can lead to different performance results~\cite{shahriari2022deep}}. 
A comparison of the results of experiments performed on different hosted ML platforms also found that out-of-the-box reproducibility is not guaranteed there~\cite{gundersen_machine_2022}. 
Another important factor is the use of GPUs, which can increase randomness compared to the use of CPUs. This is due to parallel optimization and the use of optimizers in ML frameworks, such as PyTorch and TensorFlow. As a result, some researchers have resorted to solely using CPUs for executing their experiments. However, this comes at the expense of runtime-efficiency~\cite{alahmari_challenges_2020}. 

\subsubsection{Limited access to computational resources.} 
The barrier to ML reproducibility posed by limited access to computational resources has recently become evident in the case of transformer-based Large Language Models (LLMs)~\cite{beam_challenges_2020}. 
These transformer architectures need a vast amount of data and computational resources, to which most researchers have limited access. Estimates have calculated the costs to reproduce one model to be around \$1 million to \$3.2 million~\cite{beam_challenges_2020}. Another study found that the needed computational resources are one of the most significant factors impacting reproducibility~\cite{raff_step_2019}. Especially ML models, which require computational clusters for training and optimization, are notably hard to reproduce. 

\section{Drivers for ML Reproducibility}
\label{chap:drivers}
In this section, we discuss drivers for ML reproducibility, which we subdivide into (i) technology-based drivers, (ii) procedural drivers, and (iii) drivers related to awareness and education. \rev{For every driver, we also provide case studies or examples from the literature illustrating the effectiveness of the driver for ML reproducibility.} 

\subsection{Technology-based Drivers}

\subsubsection{Hosting services.}
Utilizing hosting services offers an efficient way to share code, data, and ML model parameter settings, thus supporting the reproducibility of ML-driven research~\cite{tatman_practical_2018}. Examples of hosting services include the runtime environments of ML platforms. If the original author runs the ML experiment in such a runtime environment, e.g., Kaggle Notebooks\footnote{\url{https://www.kaggle.com/code}}, Google Colab\footnote{\url{https://colab.google/}}, or CodaLab\footnote{\url{https://codalab.org/}}, researchers attempting to reproduce the results should be able to execute the experiment within the same environment. 
The main advantage of using a hosting service is that the provider takes care of the logistics of code hosting and distribution. However, the main drawbacks are the limits on data size and computational resources. Since these hosting services are run in the cloud, there are restrictions on how many resources a single user can utilize. 
The limit on resources varies between different hosting services and is limited by users' available funds and sometimes subscription levels. Because of these limits, hosting services may not be suitable for all research purposes, especially considering the compute-intensive nature of novel ML models, such as LLMs. 

\rev{As said above, the degree to which such services offer out-of-the-box reproducibility remains highly questionable~\cite{gundersen_machine_2022}. Nevertheless, some hosting services \textit{have} effectively been used to create end-to-end reproducible AI pipelines, especially in conjunction with standardized datasets such as the National Cancer Institute Imaging Data Commons~\cite{fedorov2023national}. The effectiveness of hosting services for ML reproducibility has been shown for research in radiology~\cite{bontempi2024end} and similar reproducibility experiments have been successfully conducted in pathology research~\cite{schacherer2023nci}. As also noted in these experiments, it is important that results are reported with a quantification of the variance across different test runs, since effects of randomness could still be prevalent.} 

\subsubsection{Virtualization.}
Reproducing the environment and setup of any ML experiment requires the consideration of existing dependencies and software versions, and is usually a complex task itself. Virtualization can simplify this process by bundling the essential components of ML models and experiments, such as the dependencies and code, into a single package for sharing with other researchers.  Thus, if the authors of a paper build the experiment in a virtual environment, issues associated with setup reproduction can be greatly reduced. 
However, the adoption of virtualization by researchers depends on its user-friendliness and the effort of integration into their current workflows~\cite{boettiger_introduction_2015}. Concerns about virtualization include its limitations in allowing researchers to build upon them in a scalable manner. Traditional virtual machines (VMs) emulate an entire operating system for setting up and running experiments. The use of containerization software like Docker\footnote{\url{https://www.docker.com/}} has become more popular in recent years. Containers are more lightweight and flexible than VMs, making it easier to adapt environments for follow-up studies~\cite{boettiger_introduction_2015}. 
There are also designated platforms for computational research, such as Code Ocean\footnote{\url{https://codeocean.com/}}, that offer virtualization via so-called reproducible capsules. 
Their focus, in particular, is to simplify the virtualization process and allow researchers to focus on the research itself rather than the standardization of environments~\cite{clyburne-sherin_computational_2019}. Additionally, there are many other tools, such as ReproZip~\cite{chirigati_reprozip_2016}, one of the recommended tools by the SIGMOD Reproducibility Availability and Reproducibility Initiative\footnote{\url{https://reproducibility.sigmod.org/}} to streamline reproducibility, and DetTrace~\cite{navarro_leija_reproducible_2020}, which aims to ensure completely deterministic computations. 

\rev{The use of containers is rapidly gaining in popularity across many research fields, e.g., neuroscience and genomics~\cite{moreau2023containers}. Notably, the platform Code Ocean has been integrated into the peer reviewing process by Nature journals to support the submission process of experiments~\cite{seamless2022nature}. This widespread adoption highlights the suitability of containers to enhance experiment sharing and improve reproducibility. Furthermore, a case study has compared ten different containerization-based approaches for reproducibility~\cite{choi2023comparing}. The strengths and weaknesses of each approach were analyzed, with results demonstrating the suitability of containers to enhance reproducibility by encapsulating the computational environment and to decrease the effort for publishing reproducible ML-based experiments.}

\subsubsection{Managing sources of randomness.}
\label{section:managingrandomness}
Many different sources of randomness during ML training lead to the irreproducibility of ML research. Managing these sources of randomness, e.g., via random number seeds, deterministic algorithms, or other methods, could therefore greatly increase reproducibility. Fixed random number seeds should be used and published to make ML experiments more reproducible and control a number of sources of inherent nondeterminism. A seed is a first value used to initialize the pseudo-random number generator. When the same seed is used, the sequence of pseudo-random numbers generated is deterministic, meaning it will be the same every time the code is run. \rev{Experiments have shown that fixing random seeds can effectively ensure reproducible results when algorithms are not being executed in parallel on GPUs~\cite{ahmed_managing_2022}. Additionally, one case study has shown that achieving reproducibility for GPU-trained neural networks~\cite{chen_towards_2022} is possible through a method known as patching. Patching aims to replace non-deterministic operations with deterministic ones. In the case study, a systematic patching approach was successful in achieving reproducible image classification results for six different neural network. However, this process also leads to higher computational costs and a time overhead, which was also analyzed in the study.} Additionally, ML models should be benchmarked and evaluated with multiple random number seeds, such that the variance can be reported and inform about the true performance of an ML model~\cite{raste_quantifying_2022}. \rev{Similarly, the use of uncertainty-aware quantification metrics to evaluate ML models can also help increase reproducibility~\cite{pouchard_replicating_2020}}.

Additionally, to counteract inherent nondeterminism in reinforcement learning and achieve reproducible evaluations, there exist frameworks, such as Gym-Ignition~\cite{ferigo_gym-ignition_2020,brockman_openai_2016}, rl\_reach~\cite{aumjaud2021rl_reach}, or MinAtar~\cite{young_minatar_2019}, which act as standardized benchmarking environments. Within them, different algorithms designed for the same task can be evaluated and compared against each other in a common environment. Some frameworks counteract the effects of inherent nondeterminism by automatically controlling random seeds and evaluating algorithms over a number of runs.
Finally, it is an ongoing field of research to implement fully deterministic reinforcement learning algorithms~\cite{nagarajan_deterministic_2019} and make use of them within such frameworks~\cite{tassa2018deepmind}. 

\subsubsection{Privacy-preserving technologies.}
\label{sec:privacypreserving}
Privacy-preserving technologies support reproducibility, as they enable the collaborative training of ML models without sharing private or sensitive data. The main benefit of this is, that ML models can make use of larger and more diverse data, thus helping to decrease bias and leading to more reproducible ML models. 
The main aim of Privacy-Preserving Machine Learning (PPML) is to facilitate the use of privacy-sensitive data to create better ML models, and, to allow data owners to collaboratively train ML models on private data. 
In that regard, PPML has several requirements. First, protecting the confidentiality of the training data. Second, preventing the leakage of sensitive information from ML model parameters and outputs, i.e., to hinder the re-identification of individuals. 
Third, achieving the listed security and privacy aspects while still preserving the utility of the ML model~\cite{xu_privacy-preserving_2021}. 
To achieve this, a number of different techniques are being used and developed, mainly Differential Privacy (DP), Homomorphic Encryption (HE), Secure Multi-Party Computation (SMPC), and Federated Learning (FL)~\cite{de2021critical}. These techniques are implemented in software libraries such as TensorFlow Privacy, PySyft, ML Privacy Meter, CryptFlow, or Crypten~\cite{aslanyan_privacy-preserving_2020}. 
Furthermore, data can be made anonymous by removing identifiable personal information. However, if too much data is removed, the ML models may perform poorly. If not enough data is removed, it may still be possible to re-identify individuals by combining many different non-unique features~\cite{xu_privacy-preserving_2021}. 

An alternative approach to PPML is to generate synthetic data that captures the same information as the original data. A robust technique for creating such datasets can produce readily available datasets of nearly any size, as demonstrated in biomedical fields. This approach has led to the development of Synthea, a software package designed to generate synthetic patient data and electronic health care records~\cite{walonoski_synthea_2018}.
\rev{It is, however, important to mention that there is still a gap in efficiency between theoretical advancements and real-world applications when using PPML techniques. To this end, one study conducting reproductions of 26 state-of-the-art applications of PPML has highlighted the challenges of balancing computational efficiency, privacy guarantees, and model utility, while emphasizing the need for improved reproducibility, open-source availability, and practical scalability~\cite{khan2024wildest}.} 

\subsubsection{Tools and platforms.}
There are many tools and platforms that assist in the implementation and management of ML models and ML-based applications. 
A recent study has evaluated 19 ML tools to gain insights into their concepts constituting reproducibility support~\cite{quaranta_taxonomy_2022}. As a result, five main pillars of ML reproducibility in tools and platforms were identified: (i) code versioning, (ii) data access, (iii) data versioning, (iv) experiment logging, and (v) pipeline creation. 
Most of these pillars are associated with managing and keeping track of different artifacts created during phases of the ML lifecycle (i.e., design, development, and deployment) as for instance, datasets, labels, code, logs, environment dependencies, random number seeds, or hyperparameters~\cite{schlegel_management_2023}. 
Each of these artifacts influences the final results of the ML model. Consequently, most tools aim to collect, store, and manage these artifacts, ensuring researchers can access and use them during reproduction attempts. 
Notable are also various tools and platforms for experiment tracking~\cite{schlegel_management_2023}, such as:
\begin{itemize}
    \item \textbf{DVC\footnote{\url{https://dvc.org/}}:} A version control system for ML projects with a command-line interface similar to Git\footnote{https://git-scm.com/}. It integrates with Git, supports cloud storage, and handles large versioning of datasets. DVC ensures full code and data provenance by enabling experiment tracking. 
    \item \textbf{MLflow\footnote{\url{https://mlflow.org/}}:} An open-source tool for supporting ML experiment tracking, ML model deployment, and centralized model storage. Additionally, it provides an easy-to-use Web dashboard. 
    \item \textbf{RO-Crate\footnote{\url{https://www.researchobject.org/ro-crate/}}}: A specification, implemented by a number of tools, aimed at aggregating and describing research data and metadata~\cite{soiland-reyes_packaging_2022}. Although not specifically designed for ML, RO-Crate can aggregate and represent any resource, making it applicable for managing ML artifacts as well.
    \item \textbf{dToolAI}~\cite{hartley_dtoolai_2020}: Collects and packages ML models together with supplemental information, such as hyperparameter settings, appropriate metadata, and persistent URIs for model training data. In contrast to the other tools, dToolAI is specifically tailored towards Deep Learning models. 
\end{itemize}

AutoML platforms, such as H2O Driverless AI\footnote{\url{https://h2o.ai}}, Google Cloud AutoML\footnote{\url{https://cloud.google.com/automl}}, DataRobot\footnote{\url{https://www.datarobot.com/platform}}) are a novel subcategory of ML tools that aim to aid with every aspect of the ML lifecycle, from data aggregation to model deployment. Thus, AutoML tools could facilitate more standardized ML models and also take care of tasks like hyperparameter optimization.
It is, however, questionable how practical these tools are for reproducible ML research, since they often hide ML model optimization procedures. \rev{Recent assessments of the reproducibility of AutoML tools also came to the conclusion that current platforms cannot provide out-of-the-box-reproducibility~\cite{gundersen_machine_2022,pletzl2024reproducible}. In a qualitative analysis of the reproduction experiments, the latter study did identify areas in which such tools can be enablers for reproducibility, e.g., due to their automatic documentation capabilities. However, the authors also identified aspects that need to be addressed, such as the need for simplified tool user interfaces - as many participants were overwhelmed by tool complexity and could not make use of the documentation capabilities - and more built-in reproducibility capabilities, which support the sharing of code and data~\cite{pletzl2024reproducible}. Furthermore, some AutoML tools, such as H2O Driverless AI, aim to address problems such as model overfitting.} In the case of data leakage, this is done by checking for a strong correlation between a feature and the target and then taking action, e.g., warning the user or automatically handling it. This is, however, a very simple solution to the problem and does not address the more complex cases of data leakage that are often present in research, e.g., temporal leakage.

\subsection{Procedural Drivers}

\subsubsection{Standardized datasets and evaluation.}
Due to a lack of shared datasets, many researchers in ML-driven research - most notably in biomedical fields - have to use individually acquired data~\cite{mcdermott_reproducibility_2021}. The collection of such data is a time-consuming task and bears a significant risk of causing reproducibility issues, e.g., bias or data leakage. Often, the number of individual participants represented within datasets is not very large and, thus, findings might suffer poor generalizability.
Creating shared and standardized datasets can, therefore, (i) save researchers time in acquiring new data, (ii) facilitate the collaborative and independent maintenance and verification of data to minimize methodological errors, and (iii) support transferability and generalizability through the use of multi-institutional data~\cite{mcdermott_reproducibility_2021}. 
In addition to the standardization of datasets, data cards~\cite{pushkarna_data_2022} provide a consistent and comparable framework for reporting essential aspects of ML datasets. This includes information, e.g., about access restrictions, risks and limitations associated with the usage of the dataset, or any preprocessing steps, amongst many other contents, which are needed for reproducible ML development.

Another issue is the lack of standardized evaluation methods, which leads to reported performances of ML models often being overly optimistic~\cite{cremonesi_progress_2021}. To ensure the statistical significance of ML model evaluations, it is crucial to report performance as an aggregate of results obtained from multiple random runs, and with different random number seeds~\cite{colas_how_2018}. Furthermore, ML models should, if possible, be tested and evaluated on multiple different datasets~\cite{dror_replicability_2017}. This underscores the need for standardized evaluation methods, which can be supported by checklists or tools to prevent errors in this critical aspect of ML research. For this, similarly to data cards, model cards~\cite{mitchell_model_2019} are aimed to standardize the evaluation and reporting of the performance of ML models for a variety of use cases. Model cards should inform users about the possible applications of the ML model and its limitations.
In 2020, Google introduced the Model Card Toolkit for the creation of model cards\footnote{\url{https://research.google/blog/introducing-the-model-card-toolkit-for-easier-model-transparency-reporting/}}. In reinforcement learning, the creation of standardized evaluation pipelines is continually being researched to enable reproducible benchmarking of different reinforcement learning algorithms~\cite{khetarpal_re-evaluate_2018}. 

\rev{The National Cancer Institute Imaging Data Commons is an established example of standardized datasets in biomedical research~\cite{fedorov2023national}. As a cloud-based repository, it contains a collection of cancer imaging data and has been used successfully in reproduction experiments in combination with hosting services. Other notable examples include the MIMIC~\cite{johnson2023mimic} database for electronic health records, or OGB~\cite{hu2020open} for applying ML on graph data. 
Efforts are also being invested into increasing the reproducibility of language models, e.g., with the Holistic Evaluation of Language Models (HELM)~\cite{liang2022holistic}, which offers a broad, scenario-diverse, and multi-metric benchmarking suite for language models. As demonstrated by the authors, using HELM, new language models can be evaluated in a more comprehensive way. Furthermore, the Language Model Evaluation Harness (lm-eval) toolkit~\cite{biderman2024lessons} is a framework designed for language model evaluations and concerned with reproducibility aspects. This tool has already been used by other researchers for more reproducible language model evaluations~\cite{faysse2024croissantllm,kweon2024kormedmcqa}}. \rev{However, it is important to recognize how irreproducible most major models currently are. There exists a live-tracker of model openness\footnote{\url{https://opening-up-chatgpt.github.io/}}, which has reported that many projects, even those claiming to be open source, ``inherit undocumented data of dubious legality'', that few projects share data or model or human reinforcement learning (RLHF) weights, and that “careful scientific documentation is exceedingly rare” \cite{liesenfeld2023opening}}.

\subsubsection{Guidelines and checklists.}
There are many guidelines and checklists that outline best practices for increasing the reproducibility of ML. The guidelines are often aimed at specific parts of the ML workflow. For example, the FAIR principles\footnote{\url{https://www.go-fair.org/fair-principles/}} aim to improve the management and stewardship of scientific data by making scientific data findable, accessible, interoperable, and reusable. 
Other guidelines promote the transparency and openness of scientific reporting in general, such as the TOP guidelines\footnote{\url{https://www.cos.io/initiatives/top-guidelines}}, which target journals. 
Similarly, checklists provide a simple framework for ensuring certain criteria are met. Checklists have been applied effectively in the past, e.g., in safety-critical systems, where they were used as early as in 1935 to complete pre-flight checks in Boeing airplanes. 
A promising example is the ML checklist proposed in~\cite{pineau_improving_2021},  
which has been suggested as best practice by researchers of different fields, e.g., in chemistry~\cite{artrith_best_2021}. 
The checklist requests information about (i) the models and algorithms being used, (ii) theoretical claims in the research article, (iii) data, (iv) code, and (v) the ML experiment(s). 
However, one drawback of reproducibility checklists when used for academic conferences and journals is the additional workload they impose on already overburdened reviewers. To mitigate this, one suggestion is to leverage LLMs to assist the review process~\cite{liu_reviewergpt_2023}. 

Finally, numerous guidelines and checklists for ML reproducibility have been recommended in various research fields~\cite{artrith_best_2021}. Especially in biomedical fields, there has been a considerable adoption of guidelines and checklists, such as the TRIPOD statement~\cite{collins_transparent_2015}, the CLAIM checklist~\cite{mongan_checklist_2020}, the ROBUST-ML checklist~\cite{al-zaiti_clinicians_2022}, or PROBAST~\cite{wolff_probast_2019}. 
\rev{A systematic review in biomedical research has shown that the use of checklists is linked to increased reporting quality~\cite{han2017checklist}. The review examined 943 articles over two years and found that mandatory checklists increased the inclusion of the main methodological information needed to reproduce the experiments by 65\%.} 

\subsubsection{Model cards and model info sheets.}
Model cards~\cite{mitchell_model_2019} are documentation sheets that provide information about ML models, including their intended use, potential limitations, and ethical considerations. They aim to enhance transparency in AI, by detailing aspects such as data used for training, performance metrics, evaluation methodologies, and possible biases. Model cards help users to understand and evaluate ML models more comprehensively, such that they are not deployed in unsuited contexts, and thus to increase reproducibility. 
Similarly, model info sheets also provide documentation about ML models, but are specifically designed for the detection and prevention of data leakage in ML models~\cite{kapoor_leakage_2022}. Model info sheets are published alongside research to enable other researchers to quickly verify the validity of the data used to train ML models. They require authors to answer detailed questions about the data and corresponding train/test splits, targeting various types of data leakage~\cite{kapoor_leakage_2022}.

\rev{An empirical study investigating twelve papers, making use of ML methods for prediction, found that a third were subject to some type of data leakage~\cite{kapoor_leakage_2022}, and that in all those cases leakage errors could have been prevented by the use of model info sheets. Despite that, model info sheets have two main drawbacks: first, verifying the correctness of info sheets only works after reproducing the results; second, completing these sheets requires a certain level of expertise in ML. 
In general, model cards and model info sheets represent a promising, low-effort driver for ML reproducibility. They are especially useful in handling some of the methodological issues associated with ML models that could arise~\cite{kapoor_leakage_2022}}.

\subsection{Awareness and Education}
\label{sec:awareness}
Awareness of reproducibility issues and available training/education to support reproducibility can be a powerful driver for ML reproducibility~\cite{wiggins_replication_2019}. 

\subsubsection{Publication policies and initiatives.}
To enhance awareness and establish a minimum of reproducibility standards, the policies of scientific journals are considered an influencing factor. 
A number of journals already mandate data and/or code availability for publication~\cite{pineau_improving_2021,peng_reproducibility_2015,hardwicke_data_2018}. However, to address issues such as result manipulation, more extensive journal participation is needed to, for instance, introduce preregistration where researchers register their research intentions for future publication. This approach ensures credibility by separating the research plan from experimental outcomes~\cite{stromland_preregistration_2019,nosek_preregistration_2019}, thereby reducing spin practices, HARKing, and p-hacking~\cite{gundersen_machine_2022}. The ACM TORS (Transactions on Recommender Systems) journal exemplifies this by allowing preregistration and publishing ``reproducibility papers'' dedicated to reproduction studies and enhancing reproducibility tools. \rev{}
Apart from that, various initiatives have been launched to raise awareness of reproducibility issues. A few examples are the following:
\begin{itemize}
    \item The ReScience journal publishes peer-reviewed papers discussing attempts to reproduce original publications. These reproductions are published on GitHub\footnote{\url{https://github.com/}} and available to other researchers~\cite{rougier_sustainable_2017}.
    \item PapersWithCode.com\footnote{\url{https://paperswithcode.com/}} is a resource for (i) ML papers, accompanied by the code, (ii) datasets, and (iii) ML methods. The ML papers include a link to a repository, which features the code and other artifacts for reproducing the results. 
    \item Reproducibility challenges, where several researchers try to reproduce many recent publications in parallel, are being held frequently. These challenges allow for an analysis of the success rate of reproduction and can be used to evaluate progress over multiple years~\cite{pineau_improving_2021}. Additionally, conferences such as the European Conference on Information Retrieval (ECIR), provides special reproducibility tracks, in which researchers are encouraged to reproduce existing papers and build upon their results (e.g.,~\cite{kowald2020unfairness,muellner2021robustness,kowald2022popularity}). 
    \item The ACM has convened a new emerging interest group on reproducibility\footnote{\url{https://reproducibility.acm.org/}}. The main goals are to (i) contribute to the development of reproducibility standards, practices and policies, (ii) promote the development and evaluation of tools and methodologies, and (iii) encourage best practices. 
    \item ReproducedPapers.org is another online repository fostering reproductions. It further focuses on education by incorporating a reproduction project into a Master's level ML course at TU Delft~\cite{yildiz_reproducedpapersorg_2021}. 
\end{itemize}

\noindent \rev{As also indicated by research in information retrieval and recommender systems, increased awareness and education in the form of publication policies and initiatives can address reproducibility issues by emphasizing robust experimental practices, methodological rigor, and the development of shared resources among the different actors identified, i.e., students, educators, scholars, practitioners and decision-makers~\cite{bauer2023frontiers}.}

\section{Mapping Drivers to Barriers}
\label{chap:discussion}

\begin{figure}[!t]
	\begin{center}
		\includegraphics[width=0.90\textwidth]{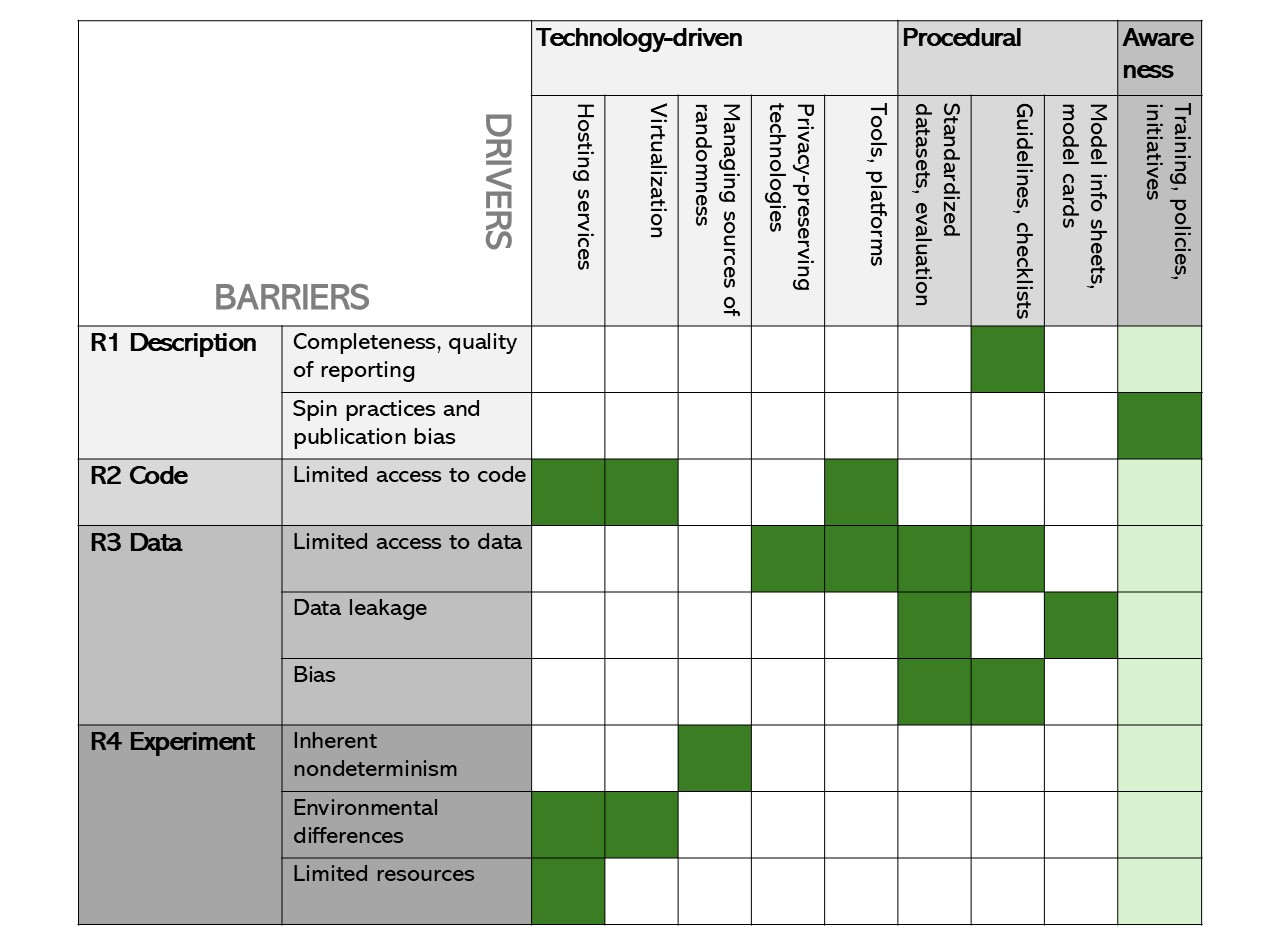} 
	\end{center}
	\caption[Drivers-Barriers-Matrix]{\rev{\textbf{Drivers-Barriers-Matrix.} We map the 9 drivers to the 9 barriers identified in this paper. The colored boxes show that a specific driver is applicable to a specific barrier. We argue that drivers related to awareness and education are, in general, applicable to address all barriers.}\vspace{-6mm}}
	\label{fig:driver_barriers_matrix}
\end{figure}

In this section, we map the drivers of reproducibility to the barriers in the form of a Drivers-Barriers Matrix. This will be based on the definition and categorization of reproducibility as a foundation (Section~\ref{s:definition}), and the identification of the major barriers (Section \ref{chap:barriers}) and drivers (Section \ref{chap:drivers}) of ML reproducibility. 
The resulting Drivers-Barriers-Matrix is depicted in Figure~\ref{fig:driver_barriers_matrix} and categorizes the barriers into the four different types of ML reproducibility, i.e., \textit{R1 Description}, \textit{R2 Code}, \textit{R3 Data}, \textit{R4 Experiment}~\cite{gundersen_fundamental_2021}, and drivers into technology-driven drivers, procedural drivers, and drivers related to awareness and education.  

Our Drivers-Barriers-Matrix shows that there are often multiple drivers for the same barrier. Consequently, there are also several possible solutions for a barrier or different aspects of a barrier. The mapping allows us to quickly assess which drivers address the different barriers and which barriers have a higher or lower number of drivers associated with them. It underlines the need for context-dependent approaches instead of ``one-size-fits-all'' solutions, as the proper selection of a suitable driver depends on the specific conditions and existing barriers relevant to any ML application. 
\rev{We describe intersections between drivers and barriers in more detail in the following and close this section with an overview of strengths and potential limitations of the identified drivers.} 

\subsubsection{R1 Description.} 
\textit{Completeness and quality of reporting}, as well as \textit{spin practices and publication bias}, present the major barriers associated with \textit{R1 Description}. These are characterized as missing information in reports and overinflated results that hinder reproducibility. 

The major drivers for \textit{completeness and report quality} are \textit{guidelines and checklists}. Guidelines provide best practices to adopt in order to achieve reproducible ML research. Furthermore, many checklists exist that comprehensively state the different pitfalls and provide information on how they can be avoided. Researchers can use them to ensure their research meets the desired standards. Furthermore, some of these checklists and guidelines are enforced by journals, such that research will only be published if certain criteria are met. 
In comparison, \textit{spin practices} are not as easily identifiable. In this case, the discussion within the research community centers around removing the incentives for inflating research results. A particularly effective driver for this is preregistration as an example for \textit{publication policies and initiatives}, where researchers submit research objectives and methods for review before conducting the research. If accepted, the research will be published regardless of the outcome (i.e., whether results are positive, negative or null), thereby minimizing spin practices.

\subsubsection{R2 Code.} 
Code sharing is essential to reproducibility, which makes \textit{limited access to code} a significant barrier in the field. However, it is often neglected as the process is not trivial. To make shared code useful to the scientific community, it is necessary to share, in addition to the source code, the information about the entire software setup and dependencies, including software versions and hardware configurations.  To assist with this, researchers may consider running their code in \textit{hosting services} or \textit{virtualization} environments, which we identified as drivers for code sharing. Both have similar advantages, i.e., they can easily be shared and made public for other researchers to use. As a consequence, it will give reproducers immediate access to code, including the complete configuration setup, such as dependencies and versions. Hosting services are a quicker and easier way of achieving this; however, they may be subject to different resource limits. \textit{Virtualization} (e.g., VMs or containers) is more difficult to set up but offers more flexibility and is not externally (e.g., by a provider) restricted in capabilities and resources. 
Furthermore, \textit{tools and platforms} can be drivers for reproducibility. A lot of ML tools provide capabilities for code versioning or other features, which are key to reproducibility. One example is dToolAI~\cite{hartley_dtoolai_2020}, which automatically logs the supplemental information of the code, i.e., metadata, hyperparameters, and more, which are essential for ML reproducibility. 

\subsubsection{R3 Data.} 
Data-related barriers are a severe obstacle to ML reproducibility due to the research fields' data-driven nature, where  \textit{limited access to data} forms a major challenge. Privacy concerns are among the crucial arguments that cause hesitation in sharing data. The need for data privacy is evident, especially in biomedical fields, which deal with patients' electronic health records. Nevertheless, it increases reproducibility issues in ML-based science and, thus, delays technological progress within these domains. However, there are several approaches that aim to meet the requirements of sharing sensitive data: \textit{Privacy-preserving technologies} allow reproducers to train ML models on private data without actually possessing the data. This way, reproduction becomes possible without violating potential privacy regulations. Other than that, the use of \textit{standardized datasets and evaluation} can support issues in regard to dataset meta-information, including the specification of train-test splits and data provenance. Once again, \textit{tools and platforms} can assist with data versioning, and numerous \textit{guidelines and checklists} have been proposed to address the provenance of data. These guidelines and checklists are designed to help researchers to avoid common pitfalls. Current initiatives are supported by journals that more frequently require data to be shared as part of a publication. 

Concerning methodological errors associated with the data, \textit{data leakage} is a major issue, which can, for instance, be mitigated using \textit{standardized datasets and evaluation}. Other drivers to solve data leakage are \textit{model info sheets and model cards}, which are provided as supplemental information to a published dataset. Even though there are some limitations to \textit{model info sheets}, they are capable of detecting all types of data leakage. 
\textit{Bias} is another methodological error, leading to irreproducible results. This is because the biased data usually does not generalize well to problems outside the experimental setup of a specific ML study. \textit{Bias} has been an important source of concern, e.g., in biomedical fields. Effects thereof can again be minimized using \textit{standardized datasets and evaluation} or specific \textit{guidelines and checklists}, e.g., ROBUST-ML~\cite{al-zaiti_clinicians_2022}. 

\begin{table}[!t]
\centering
\caption{\rev{Comparison of strengths and weaknesses of our identified drivers.}}
\label{tab:drivers}
\rev{
\begin{tabular}{|p{3cm}|p{4.5cm}|p{4.5cm}|}
\hline
\textbf{Driver}                       & \textbf{Strengths}                                                                     & \textbf{Potential Limitations}                                                       \\
\hline
\rowcolor{gray!10}
Hosting Services                      & Facilitates sharing of models, code, and datasets; increases accessibility.           & Out-of-the-box reproducibility is not yet provided and there are limits to the available compute.                  \\
\hline
\rowcolor{gray!10}
Virtualization                        & Enables environment replication (e.g., Docker, virtual machines); resolves dependency issues. & Requires technical expertise; may introduce overhead for simple experiments.         \\
\hline
\rowcolor{gray!10}
Managing sources of randomness        & Critical for deterministic outcomes; has the potential to reduce and even eliminate variances across multiple runs. & Can be hard to implement consistently across frameworks; only leads to point estimates of performance. \\
\hline
\rowcolor{gray!10}
Privacy-preserving technologies          & Expands access to sensitive datasets without compromising privacy.                    & Still an emerging field; performance trade-offs can make widespread adoption slower. \\
\hline
\rowcolor{gray!10}
Tools and platforms                   & Can streamline reproducibility practices and automatically acquire reproducibility artifacts. & Out-of-the-box reproducibility is not yet provided, and fragmentation of tools can lead to siloed solutions rather than unified workflows.   \\
\hline
\rowcolor{gray!30}
Standardized datasets and evaluation  & Provides consistency and comparability for results across studies.                   & May not generalize well to niche or domain-specific problems and can be subject to privacy concerns.                     \\
\hline
\rowcolor{gray!30}
Guidelines, checklists             & Promotes best practices through structured processes (e.g., reproducibility checklists). & Compliance can be time-consuming and may not be enforced consistently.              \\
\hline
\rowcolor{gray!30}
Model info sheets and model cards     & Improves transparency around model design and intended use.                          & Adoption is still limited; requires effort to standardize and maintain across the community. \\
\hline
\rowcolor{gray!55}
Publication policies, initiatives                    & Drive cultural change by incentivizing openness (e.g., benchmarks, competitions).     & Impact depends on community participation and is a slow process in general.                     \\
\hline
\end{tabular}
}
\end{table}

\subsubsection{R4 Experiment.} 
If an ML experiment is shared entirely and code and data are available, i.e., reproducibility type \textit{R4 Experiment}, there are still three barriers, which can lead to irreproducible results. \textit{Inherent nondeterminism} arises from the different sources of randomness in ML, and makes it difficult to achieve repeatable results, even on the same machine. There are, however, methods to \textit{manage the sources of randomness}, such as fixed random seeds and deterministic implementations, while comprehensively mitigating all sources of randomness is still a very challenging endeavor.

Another barrier is described as \textit{environmental differences}, which has two main issues associated with it, i.e., software differences and hardware differences. Both types of differences can be avoided by using either \textit{hosting services} or \textit{virtualization}; constraints can be assumed to be similar to the barrier of \textit{limited access to code}. 
\textit{Limited access to computational resources} constitutes another barrier to ML reproducibility identified in this work. The issue is particularly problematic for research using LLMs because of their need for extensive computational resources in training and reproduction. \textit{Hosting services} offer a solution, providing access to pre-trained models and allowing researchers to directly access and run respective models on-site. 
\rev{Finally, Table~\ref{tab:drivers} gives an overview of strengths and potential limitations of the identified drivers. As we can see, the choice of using a particular driver strongly depends on the given use and to what extent potential limitations are applicable for the use case.}

\section{Conclusion}
\label{chap:conclusions}
In this paper, we examined the barriers and drivers associated with the four types of ML reproducibility as outlined by Gundersen et al. (description, data, code, and experiment)~\cite{gundersen_fundamental_2021}, specifically in the cases of computer science and biomedical research. We synthesized our findings into a Drivers-Barriers-Matrix to summarize and illustrate which drivers are feasible solutions to the various barriers. We observe that the barriers to ML reproducibility can be addressed through three kinds of drivers: technology-driven solutions, procedural improvements, and enhanced awareness and education. It is important to highlight that, in theory, awareness and education can complement the other drivers and serve as a foundational basis for overcoming reproducibility-related challenges.

\rev{
One of the main issues hindering reproducibility in research appears to be rooted in the cultural aspects of research communities. As argued by \cite{chiarelli2021art,bauer2023frontiers}, the current incentives for conducting reproducible research are limited, and open research is often regarded as an unrewarded additional effort. Consequently, there is a lack of training and insufficient funding to cover the additional time and resources required by researchers. Notably, the lack of funding also impacts the ability to perform quality checks during and after the publication process. Therefore, we strongly believe that the way forward towards ML reproducibility is rooted in better education and more awareness of this topic among all involved stakeholders, e.g., students, educators, researchers, publishers, and policymakers. This, combined with the other tools and drivers described in this paper, could lead to more reproducible ML pipelines and, with this, more robust findings.} 

From a more technical perspective, the rise of AutoML tools for ML development and ML tasks performed by domain experts, (potentially) not having in-depth computer or data science knowledge, could pose another barrier to reproducibility~\cite{Haberl2025}.
\rev{We thus believe future work should address the increasing use of AutoML tools for AI development in research (\cite{Haberl2025}) among non-computer or data science experts. While these easy-to-use tools can standardize ML workflows by default and include documentation features, domain experts often lack the necessary expertise to recognize potential problems associated with ML, such as biased or imbalanced data. Thus, research on reproducibility should emphasize this challenge and aim to establish standards and guidelines for the use of No and Low Code ML tools in research, as well as the training required for their responsible application.}

In summary, we hope that our paper provides practical guidance and orientation for researchers employing ML and clarifies the current state of play. Of course, in such a dynamic and fast-paced research area, this discussion opens up a series of further questions and avenues for exploration. We recommend further investigation of the various issues and potential solutions laid out here. We would also encourage further investigation into the potential role of platforms~\cite{gundersen_machine_2022} or foundation models~\cite{hosseini_open_2024} in further exacerbating or alleviating these challenges. 

\textbf{Acknowledgements.} This research is supported by the Horizon Europe project TIER2 (GA: 101094817), and the FFG COMET program. 

\bibliographystyle{splncs04}
\bibliography{reproml.bib}

\begin{thebibliography}{100}
\providecommand{\url}[1]{\texttt{#1}}
\providecommand{\urlprefix}{URL }
\providecommand{\doi}[1]{https://doi.org/#1}

\bibitem{ahmed_managing_2022}
Ahmed, H., Lofstead, J.: Managing {Randomness} to {Enable} {Reproducible} {Machine} {Learning}. In: Proceedings of the 5th {International} {Workshop} on {Practical} {Reproducible} {Evaluation} of {Computer} {Systems}. pp. 15--20. ACM, Minneapolis MN USA (2022)

\bibitem{ahmed_measuring_2022}
Ahmed, H., Tchoua, R., Lofstead, J.: Measuring {Reproduciblity} of {Machine} {Learning} {Methods} for {Medical} {Diagnosis}. In: 2022 {Fourth} {International} {Conference} on {Transdisciplinary} {AI} ({TransAI}). pp. 9--16 (2022)

\bibitem{al-zaiti_clinicians_2022}
Al-Zaiti, S.S., Alghwiri, A.A., Hu, X., Clermont, G., Peace, A., Macfarlane, P., Bond, R.: A clinician’s guide to understanding and critically appraising machine learning studies: a checklist for {Ruling} {Out} {Bias} {Using} {Standard} {Tools} in {Machine} {Learning} ({ROBUST}-{ML}). European Heart Journal - Digital Health  \textbf{3}(2),  125--140 (2022)

\bibitem{alahmari_challenges_2020}
Alahmari, S.S., Goldgof, D.B., Mouton, P.R., Hall, L.O.: Challenges for the {Repeatability} of {Deep} {Learning} {Models}. IEEE Access  \textbf{8},  211860--211868 (2020), conference Name: IEEE Access

\bibitem{albertoni_reproducibility_2023}
Albertoni, R., Colantonio, S., Skrzypczy{\'n}ski, P., Stefanowski, J.: Reproducibility of machine learning: Terminology, recommendations and open issues. arXiv preprint arXiv:2302.12691  (2023)

\bibitem{andaur_navarro_systematic_2023}
Andaur~Navarro, C.L., Damen, J.A., Takada, T., Nijman, S.W., Dhiman, P., Ma, J., Collins, G.S., Bajpai, R., Riley, R.D., Moons, K.G., Hooft, L.: Systematic review finds “spin” practices and poor reporting standards in studies on machine learning-based prediction models. Journal of Clinical Epidemiology  \textbf{158},  99--110 (2023)

\bibitem{artrith_best_2021}
Artrith, N., Butler, K.T., Coudert, F.X., Han, S., Isayev, O., Jain, A., Walsh, A.: Best practices in machine learning for chemistry. Nature Chemistry  \textbf{13}(6),  505--508 (2021)

\bibitem{aslanyan_privacy-preserving_2020}
Aslanyan, Z., Vasilikos, P.: Privacy-{Preserving} {Machine} {Learning} (2020)

\bibitem{aumjaud2021rl_reach}
Aumjaud, P., McAuliffe, D., Lera, F.J.R., Cardiff, P.: rl\_reach: Reproducible reinforcement learning experiments for robotic reaching tasks. Software Impacts  \textbf{8},  100061 (2021)

\bibitem{baker_1500_2016}
Baker, M.: 1,500 scientists lift the lid on reproducibility. Nature  \textbf{533}(7604),  452--454 (2016)

\bibitem{banks2016evidence}
Banks, G.C., Rogelberg, S.G., Woznyj, H.M., Landis, R.S., Rupp, D.E.: Evidence on questionable research practices: The good, the bad, and the ugly. Journal of Business and Psychology  \textbf{31},  323--338 (2016)

\bibitem{bauer2023frontiers}
Bauer, C., Carterette, B., Ferro, N., Fuhr, N., Faggioli, G.: Frontiers of information access experimentation for research and education (dagstuhl seminar 23031). In: Dagstuhl Reports. vol.~13. Schloss Dagstuhl-Leibniz-Zentrum f{\"u}r Informatik (2023)

\bibitem{beam_challenges_2020}
Beam, A.L., Manrai, A.K., Ghassemi, M.: Challenges to the {Reproducibility} of {Machine} {Learning} {Models} in {Health} {Care}. JAMA  \textbf{323}(4),  305--306 (2020)

\bibitem{belz_systematic_2021}
Belz, A., Agarwal, S., Shimorina, A., Reiter, E.: A systematic review of reproducibility research in natural language processing. In: Proceedings of the 16th Conference of the European Chapter of the Association for Computational Linguistics. pp. 381--393 (01 2021)

\bibitem{biagioli2020gaming}
Biagioli, M., Lippman, A.: Gaming the metrics: Misconduct and manipulation in academic research. Mit Press (2020)

\bibitem{biderman2024lessons}
Biderman, S., Schoelkopf, H., Sutawika, L., Gao, L., Tow, J., Abbasi, B., Aji, A.F., Ammanamanchi, P.S., Black, S., Clive, J., et~al.: Lessons from the trenches on reproducible evaluation of language models. arXiv preprint arXiv:2405.14782  (2024)

\bibitem{boettiger_introduction_2015}
Boettiger, C.: An introduction to {Docker} for reproducible research. ACM SIGOPS Operating Systems Review  \textbf{49}(1),  71--79 (2015)

\bibitem{bontempi2024end}
Bontempi, D., Nuernberg, L., Pai, S., Krishnaswamy, D., Thiriveedhi, V., Hosny, A., Mak, R.H., Farahani, K., Kikinis, R., Fedorov, A., et~al.: End-to-end reproducible ai pipelines in radiology using the cloud. Nature Communications  \textbf{15}(1), ~6931 (2024)

\bibitem{brockman_openai_2016}
Brockman, G., Cheung, V., Pettersson, L., Schneider, J., Schulman, J., Tang, J., Zaremba, W.: Openai gym. arXiv preprint arXiv:1606.01540  (2016)

\bibitem{burrows2012living}
Burrows, R.: Living with the h-index? metric assemblages in the contemporary academy. The sociological review  \textbf{60}(2),  355--372 (2012)

\bibitem{chen_towards_2022}
Chen, B., Wen, M., Shi, Y., Lin, D., Rajbahadur, G.K., Jiang, Z.M.J.: Towards training reproducible deep learning models. In: Proceedings of the 44th International Conference on Software Engineering. p. 2202–2214. ICSE '22, ACM, New York, NY, USA (2022)

\bibitem{chiarelli2021art}
Chiarelli, A., Loffreda, L., Johnson, R.: The Art of Publishing Reproducible Research Outputs: Supporting emerging practices through cultural and technological innovation. Knowledge Exchange (2021)

\bibitem{chirigati_reprozip_2016}
Chirigati, F., Rampin, R., Shasha, D., Freire, J.: {ReproZip}: {Computational} {Reproducibility} {With} {Ease}. In: Proceedings of the 2016 {International} {Conference} on {Management} of {Data}. pp. 2085--2088. {SIGMOD} '16, ACM, New York, NY, USA (2016)

\bibitem{choi2023comparing}
Choi, Y.D., Roy, B., Nguyen, J., Ahmad, R., Maghami, I., Nassar, A., Li, Z., Castronova, A.M., Malik, T., Wang, S., et~al.: Comparing containerization-based approaches for reproducible computational modeling of environmental systems. Environmental Modelling \& Software  \textbf{167},  105760 (2023)

\bibitem{clyburne-sherin_computational_2019}
Clyburne-Sherin, A., Fei, X., Green, S.A.: Computational {Reproducibility} via {Containers} in {Psychology}. Meta-Psychology  \textbf{3} (2019)

\bibitem{colas_how_2018}
Colas, C., Sigaud, O., Oudeyer, P.Y.: How many random seeds? statistical power analysis in deep reinforcement learning experiments. arXiv preprint arXiv:1806.08295  (2018)

\bibitem{open2015estimating}
Collaboration, O.S.: Estimating the reproducibility of psychological science. Science  \textbf{349}(6251),  aac4716 (2015)

\bibitem{collins_transparent_2015}
Collins, G.S., Reitsma, J.B., Altman, D.G., Moons, K.G.: Transparent {Reporting} of a {Multivariable} {Prediction} {Model} for {Individual} {Prognosis} or {Diagnosis} ({TRIPOD}). Circulation  \textbf{131}(2),  211--219 (2015), publisher: American Heart Association

\bibitem{cremonesi_progress_2021}
Cremonesi, P., Jannach, D.: Progress in {Recommender} {Systems} {Research}: {Crisis}? {What} {Crisis}? AI Magazine  \textbf{42}(3),  43--54 (2021)

\bibitem{de2021critical}
De~Cristofaro, E.: A critical overview of privacy in machine learning. IEEE Security \& Privacy  \textbf{19}(4),  19--27 (2021)

\bibitem{dror_replicability_2017}
Dror, R., Baumer, G., Bogomolov, M., Reichart, R.: Replicability {Analysis} for {Natural} {Language} {Processing}: {Testing} {Significance} with {Multiple} {Datasets}. Transactions of the Association for Computational Linguistics  \textbf{5},  471--486 (2017)

\bibitem{dwivedi2021artificial}
Dwivedi, Y.K., Hughes, L., Ismagilova, E., Aarts, G., Coombs, C., Crick, T., Duan, Y., Dwivedi, R., Edwards, J., Eirug, A., et~al.: Artificial intelligence (ai): Multidisciplinary perspectives on emerging challenges, opportunities, and agenda for research, practice and policy. International journal of information management  \textbf{57},  101994 (2021)

\bibitem{d2024towards}
D’Antonoli, T.A., Cuocolo, R., Baessler, B., Dos~Santos, D.P.: Towards reproducible radiomics research: introduction of a database for radiomics studies. European Radiology  \textbf{34}(1), ~436 (2024)

\bibitem{seamless2022nature}
Editorial, N.C.S.: Seamless sharing and peer review of code. Nature Computational Science  \textbf{2}(12), ~773 (2022)

\bibitem{errington2021investigating}
Errington, T.M., Mathur, M., Soderberg, C.K., Denis, A., Perfito, N., Iorns, E., Nosek, B.A.: Investigating the replicability of preclinical cancer biology. Elife  \textbf{10},  e71601 (2021)

\bibitem{faysse2024croissantllm}
Faysse, M., Fernandes, P., Guerreiro, N., Loison, A., Alves, D., Corro, C., Boizard, N., Alves, J., Rei, R., Martins, P., et~al.: Croissantllm: A truly bilingual french-english language model. arXiv preprint arXiv:2402.00786  (2024)

\bibitem{fedorov2023national}
Fedorov, A., Longabaugh, W.J., Pot, D., Clunie, D.A., Pieper, S.D., Gibbs, D.L., Bridge, C., Herrmann, M.D., Homeyer, A., Lewis, R., et~al.: National cancer institute imaging data commons: toward transparency, reproducibility, and scalability in imaging artificial intelligence. Radiographics  \textbf{43}(12),  e230180 (2023)

\bibitem{ferigo_gym-ignition_2020}
Ferigo, D., Traversaro, S., Metta, G., Pucci, D.: Gym-{Ignition}: {Reproducible} {Robotic} {Simulations} for {Reinforcement} {Learning}. In: 2020 {IEEE}/{SICE} {International} {Symposium} on {System} {Integration} ({SII}). pp. 885--890 (2020)

\bibitem{fidler_reproducibility_2021}
Fidler, F., Wilcox, J.: Reproducibility of {Scientific} {Results}. In: Zalta, E.N. (ed.) The {Stanford} {Encyclopedia} of {Philosophy}. Metaphysics Research Lab, Stanford University, summer 2021 edn. (2021)

\bibitem{freedman2015economics}
Freedman, L.P., Cockburn, I.M., Simcoe, T.S.: The economics of reproducibility in preclinical research. PLoS biology  \textbf{13}(6),  e1002165 (2015)

\bibitem{gibney_could_2022}
Gibney, E.: Could machine learning fuel a reproducibility crisis in science? Nature  \textbf{608}(7922),  250--251 (2022), bandiera\_abtest: a Cg\_type: News Number: 7922 Publisher: Nature Publishing Group Subject\_term: Machine learning, Publishing, Mathematics and computing

\bibitem{goodman_what_2016}
Goodman, S.N., Fanelli, D., Ioannidis, J.P.: What does research reproducibility mean? Science translational medicine  \textbf{8}(341),  341ps12--341ps12 (2016)

\bibitem{gundersen_fundamental_2021}
Gundersen, O.E.: The fundamental principles of reproducibility. Philosophical Transactions of the Royal Society A: Mathematical, Physical and Engineering Sciences  \textbf{379}(2197),  20200210 (2021), publisher: Royal Society

\bibitem{gundersen2023sources}
Gundersen, O.E., Coakley, K., Kirkpatrick, C., Gil, Y.: Sources of irreproducibility in machine learning: A review. arXiv preprint arXiv:2204.07610  (2023)

\bibitem{gundersen_state_2018}
Gundersen, O.E., Kjensmo, S.: State of the {Art}: {Reproducibility} in {Artificial} {Intelligence}. Proceedings of the AAAI Conference on Artificial Intelligence  \textbf{32}(1) (2018)

\bibitem{gundersen_machine_2022}
Gundersen, O.E., Shamsaliei, S., Isdahl, R.J.: Do machine learning platforms provide out-of-the-box reproducibility? Future Generation Computer Systems  \textbf{126},  34--47 (2022)

\bibitem{gundersen2023reporting}
Gundersen, O.E., Shamsaliei, S., Kj{\ae}rnli, H.S., Langseth, H.: On reporting robust and trustworthy conclusions from model comparison studies involving neural networks and randomness. In: Proceedings of the 2023 ACM Conference on Reproducibility and Replicability. pp. 37--61 (2023)

\bibitem{Haberl2025}
Haberl, A.;~Thalmann, S.: Automated machine learning in research – a literature review. In: Proceedings of the 58th Hawaii International Conference on System Sciences (2025)

\bibitem{han2017checklist}
Han, S., Olonisakin, T.F., Pribis, J.P., Zupetic, J., Yoon, J.H., Holleran, K.M., Jeong, K., Shaikh, N., Rubio, D.M., Lee, J.S.: A checklist is associated with increased quality of reporting preclinical biomedical research: a systematic review. PloS one  \textbf{12}(9),  e0183591 (2017)

\bibitem{hardwicke_data_2018}
Hardwicke, T.E., Mathur, M.B., MacDonald, K., Nilsonne, G., Banks, G.C., Kidwell, M.C., Hofelich~Mohr, A., Clayton, E., Yoon, E.J., Henry~Tessler, M., Lenne, R.L., Altman, S., Long, B., Frank, M.C.: Data availability, reusability, and analytic reproducibility: evaluating the impact of a mandatory open data policy at the journal {Cognition}. Royal Society Open Science  \textbf{5}(8),  180448 (2018), publisher: Royal Society

\bibitem{hardwicke2020empirical}
Hardwicke, T.E., Wallach, J.D., Kidwell, M.C., Bendixen, T., Cr{\"u}well, S., Ioannidis, J.P.: An empirical assessment of transparency and reproducibility-related research practices in the social sciences (2014--2017). Royal Society open science  \textbf{7}(2),  190806 (2020)

\bibitem{hartley_dtoolai_2020}
Hartley, M., Olsson, T.S.G.: {dtoolAI}: {Reproducibility} for {Deep} {Learning}. Patterns  \textbf{1}(5),  100073 (2020)

\bibitem{heil_reproducibility_2021}
Heil, B.J., Hoffman, M.M., Markowetz, F., Lee, S.I., Greene, C.S., Hicks, S.C.: Reproducibility standards for machine learning in the life sciences. Nature Methods  \textbf{18}(10),  1132--1135 (2021)

\bibitem{hong_evaluation_2013}
Hong, S.Y., Koo, M.S., Jang, J., Kim, J.E.E., Park, H., Joh, M.S., Kang, J.H., Oh, T.J.: An {Evaluation} of the {Software} {System} {Dependency} of a {Global} {Atmospheric} {Model}. Monthly Weather Review  \textbf{141}(11),  4165--4172 (2013), publisher: American Meteorological Society Section: Monthly Weather Review

\bibitem{hosseini_open_2024}
Hosseini, M., Horbach, S.P., Holmes, K., Ross-Hellauer, T.: Open science at the generative ai turn: An exploratory analysis of challenges and opportunities. Quantitative Science Studies pp. 1--24 (2024)

\bibitem{hu2020open}
Hu, W., Fey, M., Zitnik, M., Dong, Y., Ren, H., Liu, B., Catasta, M., Leskovec, J.: Open graph benchmark: Datasets for machine learning on graphs. Advances in neural information processing systems  \textbf{33},  22118--22133 (2020)

\bibitem{hutson_artificial_2018}
Hutson, M.: Artificial intelligence faces reproducibility crisis {Unpublished} code and sensitivity to training conditions make many claims hard to verify. Science  \textbf{359}(6377),  725--726 (2018)

\bibitem{hutson_missing_2018}
Hutson, M.: Missing data hinder replication of artificial intelligence studies. Science  (2018)

\bibitem{ioannidis2005most}
Ioannidis, J.P.: Why most published research findings are false. PLoS medicine  \textbf{2}(8), ~e124 (2005)

\bibitem{iqbal2016reproducible}
Iqbal, S.A., Wallach, J.D., Khoury, M.J., Schully, S.D., Ioannidis, J.P.: Reproducible research practices and transparency across the biomedical literature. PLoS biology  \textbf{14}(1),  e1002333 (2016)

\bibitem{ivanescu_importance_2016}
Ivanescu, A.E., Li, P., George, B., Brown, A.W., Keith, S.W., Raju, D., Allison, D.B.: The {Importance} of {Prediction} {Model} {Validation} and {Assessment} in {Obesity} and {Nutrition} {Research}. International journal of obesity (2005)  \textbf{40}(6),  887--894 (2016)

\bibitem{johnson2023mimic}
Johnson, A.E., Bulgarelli, L., Shen, L., Gayles, A., Shammout, A., Horng, S., Pollard, T.J., Hao, S., Moody, B., Gow, B., et~al.: Mimic-iv, a freely accessible electronic health record dataset. Scientific data  \textbf{10}(1), ~1 (2023)

\bibitem{kamel_rahimi_machine_2022}
Kamel~Rahimi, A., Canfell, O.J., Chan, W., Sly, B., Pole, J.D., Sullivan, C., Shrapnel, S.: Machine learning models for diabetes management in acute care using electronic medical records: {A} systematic review. International Journal of Medical Informatics  \textbf{162},  104758 (2022)

\bibitem{kapoor_leakage_2022}
Kapoor, S., Narayanan, A.: Leakage and the reproducibility crisis in machine-learning-based science. Patterns  \textbf{4}(9),  100804 (2023)

\bibitem{khan2024wildest}
Khan, T., Budzys, M., Nguyen, K., Michalas, A.: Wildest dreams: Reproducible research in privacy-preserving neural network training. arXiv preprint arXiv:2403.03592  (2024)

\bibitem{khetarpal_re-evaluate_2018}
Khetarpal, K., Ahmed, Z., Cianflone, A., Islam, R., Pineau, J.: {RE}-{EVALUATE}: {Reproducibility} in {Evaluating} {Reinforcement} {Learning} {Algorithms}. In: 2nd Reproducibility in Machine Learning Workshop at ICML (2018)

\bibitem{koenigstorfer_black_2024}
Koenigstorfer, F., Haberl, A., Kowald, D., Ross-Hellauer, T., Thalmann, S.: Black box or open science? assessing reproducibility-related documentation in ai research. In: Proceedings of the 57th Hawaii International Conference on System Sciences (2024)

\bibitem{kowald2022popularity}
Kowald, D., Lacic, E.: Popularity bias in collaborative filtering-based multimedia recommender systems. In: BIAS WS at ECIR. pp. 1--11. Springer (2022)

\bibitem{kowald2020unfairness}
Kowald, D., Schedl, M., Lex, E.: The unfairness of popularity bias in music recommendation: A reproducibility study. In: 42nd European Conference on IR Research, ECIR 2020. pp. 35--42. Springer (2020)

\bibitem{kowald2024establishing}
Kowald, D., Scher, S., Pammer-Schindler, V., M{\"u}llner, P., Waxnegger, K., Demelius, L., Fessl, A., Toller, M., Mendoza~Estrada, I.G., {\v{S}}imi{\'c}, I., et~al.: Establishing and evaluating trustworthy ai: overview and research challenges. Frontiers in Big Data  \textbf{7},  1467222 (2024)

\bibitem{kweon2024kormedmcqa}
Kweon, S., Choi, B., Kim, M., Park, R.W., Choi, E.: Kormedmcqa: Multi-choice question answering benchmark for korean healthcare professional licensing examinations. arXiv preprint arXiv:2403.01469  (2024)

\bibitem{lariviere2019journal}
Lariviere, V., Sugimoto, C.R.: The journal impact factor: A brief history, critique, and discussion of adverse effects. Springer handbook of science and technology indicators pp. 3--24 (2019)

\bibitem{lee2013bias}
Lee, C.J., Sugimoto, C.R., Zhang, G., Cronin, B.: Bias in peer review. Journal of the American Society for information Science and Technology  \textbf{64}(1),  2--17 (2013)

\bibitem{leventi-peetz_deep_2022}
Leventi-Peetz, A.M., {\"O}streich, T.: Deep learning reproducibility and explainable ai (xai). arXiv preprint arXiv:2202.11452  (2022)

\bibitem{liang2022holistic}
Liang, P., Bommasani, R., Lee, T., Tsipras, D., Soylu, D., Yasunaga, M., Zhang, Y., Narayanan, D., Wu, Y., Kumar, A., et~al.: Holistic evaluation of language models. arXiv preprint arXiv:2211.09110  (2022)

\bibitem{liesenfeld2023opening}
Liesenfeld, A., Lopez, A., Dingemanse, M.: Opening up chatgpt: Tracking openness, transparency, and accountability in instruction-tuned text generators. In: Proceedings of the 5th international conference on conversational user interfaces. pp.~1--6 (2023)

\bibitem{liu_reviewergpt_2023}
Liu, R., Shah, N.B.: Reviewergpt? an exploratory study on using large language models for paper reviewing. arXiv preprint arXiv:2306.00622  (2023)

\bibitem{mcdermott_reproducibility_2021}
McDermott, M.B.A., Wang, S., Marinsek, N., Ranganath, R., Foschini, L., Ghassemi, M.: Reproducibility in machine learning for health research: {Still} a ways to go. Science Translational Medicine  \textbf{13}(586),  eabb1655 (2021)

\bibitem{mehrabi2021survey}
Mehrabi, N., Morstatter, F., Saxena, N., Lerman, K., Galstyan, A.: A survey on bias and fairness in machine learning. ACM computing surveys (CSUR)  \textbf{54}(6),  1--35 (2021)

\bibitem{mitchell_model_2019}
Mitchell, M., Wu, S., Zaldivar, A., Barnes, P., Vasserman, L., Hutchinson, B., Spitzer, E., Raji, I.D., Gebru, T.: Model {Cards} for {Model} {Reporting}. In: Proceedings of the {Conference} on {Fairness}, {Accountability}, and {Transparency}. pp. 220--229. {FAT}* '19, ACM, New York, NY, USA (2019)

\bibitem{mongan_checklist_2020}
Mongan, J., Moy, L., Kahn, C.E.: Checklist for {Artificial} {Intelligence} in {Medical} {Imaging} ({CLAIM}): {A} {Guide} for {Authors} and {Reviewers}. Radiology: Artificial Intelligence  \textbf{2}(2),  e200029 (2020), publisher: Radiological Society of North America

\bibitem{moreau2023containers}
Moreau, D., Wiebels, K., Boettiger, C.: Containers for computational reproducibility. Nature Reviews Methods Primers  \textbf{3}(1), ~50 (2023)

\bibitem{muellner2021robustness}
Muellner, P., Kowald, D., Lex, E.: Robustness of meta matrix factorization against strict privacy constraints. In: 43rd European Conference on IR Research, ECIR 2021. pp. 107--119. Springer (2021)

\bibitem{munafo_manifesto_2017}
Munafò, M.R., Nosek, B.A., Bishop, D.V.M., Button, K.S., Chambers, C.D., Percie~du Sert, N., Simonsohn, U., Wagenmakers, E.J., Ware, J.J., Ioannidis, J.P.A.: A manifesto for reproducible science. Nature Human Behaviour  \textbf{1}(1), ~1--9 (2017), publisher: Nature Publishing Group

\bibitem{nagarajan_deterministic_2019}
Nagarajan, P., Warnell, G., Stone, P.: Deterministic implementations for reproducibility in deep reinforcement learning. arXiv preprint arXiv:1809.05676  (2018)

\bibitem{navarro_leija_reproducible_2020}
Navarro~Leija, O.S., Shiptoski, K., Scott, R.G., Wang, B., Renner, N., Newton, R.R., Devietti, J.: Reproducible {Containers}. In: Proceedings of the {Twenty}-{Fifth} {International} {Conference} on {Architectural} {Support} for {Programming} {Languages} and {Operating} {Systems}. pp. 167--182. ACM, Lausanne Switzerland (2020)

\bibitem{nosek_preregistration_2019}
Nosek, B.A., Beck, E.D., Campbell, L., Flake, J.K., Hardwicke, T.E., Mellor, D.T., van~’t Veer, A.E., Vazire, S.: Preregistration {Is} {Hard}, {And} {Worthwhile}. Trends in Cognitive Sciences  \textbf{23}(10),  815--818 (2019)

\bibitem{nosek2022replicability}
Nosek, B.A., Hardwicke, T.E., Moshontz, H., Allard, A., Corker, K.S., Dreber, A., Fidler, F., Hilgard, J., Kline~Struhl, M., Nuijten, M.B., et~al.: Replicability, robustness, and reproducibility in psychological science. Annual review of psychology  \textbf{73}(1),  719--748 (2022)

\bibitem{ooi2023potential}
Ooi, K.B., Tan, G.W.H., Al-Emran, M., Al-Sharafi, M.A., Capatina, A., Chakraborty, A., Dwivedi, Y.K., Huang, T.L., Kar, A.K., Lee, V.H., et~al.: The potential of generative artificial intelligence across disciplines: Perspectives and future directions. Journal of Computer Information Systems pp. 1--32 (2023)

\bibitem{panaretos_comparison_2018}
Panaretos, D., Koloverou, E., Dimopoulos, A.C., Kouli, G.M., Vamvakari, M., Tzavelas, G., Pitsavos, C., Panagiotakos, D.B.: A comparison of statistical and machine-learning techniques in evaluating the association between dietary patterns and 10-year cardiometabolic risk (2002–2012): the {ATTICA} study. British Journal of Nutrition  \textbf{120}(3),  326--334 (2018), publisher: Cambridge University Press

\bibitem{paullada_data_2021}
Paullada, A., Raji, I.D., Bender, E.M., Denton, E., Hanna, A.: Data and its (dis)contents: {A} survey of dataset development and use in machine learning research. Patterns  \textbf{2}(11),  100336 (2021)

\bibitem{peng_reproducibility_2015}
Peng, R.: The reproducibility crisis in science: {A} statistical counterattack. Significance  \textbf{12}(3),  30--32 (2015)

\bibitem{pineau_improving_2021}
Pineau, J., Vincent-Lamarre, P., Sinha, K., Larivière, V., Beygelzimer, A., d'Alché Buc, F., Fox, E., Larochelle, H.: Improving reproducibility in machine learning research (a report from the {NeurIPS} 2019 reproducibility program). The Journal of Machine Learning Research  \textbf{22}(1),  164:7459--164:7478 (2021)

\bibitem{pletzl2024reproducible}
Pletzl, S., Haberl, A., Ross-Hellauer, T., Thalmann, S.: Reproducible automl: An assessment of research reproducibility of no-code automl tools. In: Wirtschaftsinformatik 2024 Proceedings (2024)

\bibitem{pontika2022indicators}
Pontika, N., Klebel, T., Correia, A., Metzler, H., Knoth, P., Ross-Hellauer, T.: Indicators of research quality, quantity, openness, and responsibility in institutional review, promotion, and tenure policies across seven countries. Quantitative Science Studies  \textbf{3}(4),  888--911 (2022)

\bibitem{pouchard_replicating_2020}
Pouchard, L., Lin, Y., Van~Dam, H.: Replicating {Machine} {Learning} {Experiments} in {Materials} {Science}. In: Parallel {Computing}: {Technology} {Trends}, pp. 743--755. IOS Press (2020)

\bibitem{provenzano_radiologist_2021}
Provenzano, D., Rao, Y.J., Goyal, S., Haji-Momenian, S., Lichtenberger, J., Loew, M.: Radiologist vs {Machine} {Learning}: {A} {Comparison} of {Performance} in {Cancer} {Imaging}. In: 2021 {IEEE} {Applied} {Imagery} {Pattern} {Recognition} {Workshop} ({AIPR}). pp. 1--10 (2021), iSSN: 2332-5615

\bibitem{pushkarna_data_2022}
Pushkarna, M., Zaldivar, A., Kjartansson, O.: Data {Cards}: {Purposeful} and {Transparent} {Dataset} {Documentation} for {Responsible} {AI}. In: Proceedings of the 2022 {ACM} {Conference} on {Fairness}, {Accountability}, and {Transparency}. pp. 1776--1826. {FAccT} '22, ACM, New York, NY, USA (2022)

\bibitem{quaranta_taxonomy_2022}
Quaranta, L., Calefato, F., Lanubile, F.: A {Taxonomy} of {Tools} for {Reproducible} {Machine} {Learning} {Experiments}. CEUR Workshop Proceedings  (2022)

\bibitem{raff_step_2019}
Raff, E.: A {Step} {Toward} {Quantifying} {Independently} {Reproducible} {Machine} {Learning} {Research}. In: Advances in {Neural} {Information} {Processing} {Systems}. vol.~32. Curran Associates, Inc. (2019)

\bibitem{raste_quantifying_2022}
Raste, S., Singh, R., Vaughan, J., Nair, V.N.: Quantifying {Inherent} {Randomness} in {Machine} {Learning} {Algorithms}. SSRN Electronic Journal  (2022)

\bibitem{rijcke2016evaluation}
Rijcke, S.d., Wouters, P.F., Rushforth, A.D., Franssen, T.P., Hammarfelt, B.: Evaluation practices and effects of indicator use—a literature review. Research evaluation  \textbf{25}(2),  161--169 (2016)

\bibitem{rougier_sustainable_2017}
Rougier, N.P., Hinsen, K., Alexandre, F., Arildsen, T., Barba, L.A., Benureau, F.C.Y., Brown, C.T., Buyl, P.d., Caglayan, O., Davison, A.P., Delsuc, M.A., Detorakis, G., Diem, A.K., Drix, D., Enel, P., Girard, B., Guest, O., Hall, M.G., Henriques, R.N., Hinaut, X., Jaron, K.S., Khamassi, M., Klein, A., Manninen, T., Marchesi, P., McGlinn, D., Metzner, C., Petchey, O., Plesser, H.E., Poisot, T., Ram, K., Ram, Y., Roesch, E., Rossant, C., Rostami, V., Shifman, A., Stachelek, J., Stimberg, M., Stollmeier, F., Vaggi, F., Viejo, G., Vitay, J., Vostinar, A.E., Yurchak, R., Zito, T.: Sustainable computational science: the {ReScience} initiative. PeerJ Computer Science  \textbf{3}, ~e142 (2017), publisher: PeerJ Inc.

\bibitem{rouzrokh_mitigating_2022}
Rouzrokh, P., Khosravi, B., Faghani, S., Moassefi, M., Vera~Garcia, D.V., Singh, Y., Zhang, K., Conte, G.M., Erickson, B.J.: Mitigating {Bias} in {Radiology} {Machine} {Learning}: 1. {Data} {Handling}. Radiology: Artificial Intelligence  \textbf{4}(5),  e210290 (2022)

\bibitem{saidi2024unravelingoveroptimismpublicationbias}
Saidi, P., Dasarathy, G., Berisha, V.: Unraveling overoptimism and publication bias in ml-driven science (2024), \url{https://arxiv.org/abs/2405.14422}

\bibitem{schacherer2023nci}
Schacherer, D.P., Herrmann, M.D., Clunie, D.A., H{\"o}fener, H., Clifford, W., Longabaugh, W.J., Pieper, S., Kikinis, R., Fedorov, A., Homeyer, A.: The nci imaging data commons as a platform for reproducible research in computational pathology. Computer methods and programs in biomedicine  \textbf{242},  107839 (2023)

\bibitem{schlegel_management_2023}
Schlegel, M., Sattler, K.U.: Management of {Machine} {Learning} {Lifecycle} {Artifacts}: {A} {Survey}. ACM SIGMOD Record  \textbf{51}(4),  18--35 (2023)

\bibitem{semmelrock2023reproducibility}
Semmelrock, H., Kopeinik, S., Theiler, D., Ross-Hellauer, T., Kowald, D.: Reproducibility in machine learning-driven research. arXiv preprint arXiv:2307.10320  (2023)

\bibitem{shahriari2022deep}
Shahriari, M., Ramler, R., Fischer, L.: How do deep-learning framework versions affect the reproducibility of neural network models? Machine Learning and Knowledge Extraction  \textbf{4}(4),  888--911 (2022)

\bibitem{smaldino2016natural}
Smaldino, P.E., McElreath, R.: The natural selection of bad science. Royal Society open science  \textbf{3}(9),  160384 (2016)

\bibitem{soiland-reyes_packaging_2022}
Soiland-Reyes, S., Sefton, P., Crosas, M., Castro, L.J., Coppens, F., Fernández, J.M., Garijo, D., Grüning, B., La~Rosa, M., Leo, S., Ó~Carragáin, E., Portier, M., Trisovic, A., Community, R.C., Groth, P., Goble, C.: Packaging research artefacts with {RO}-{Crate}. Data Science  \textbf{5}(2),  97--138 (2022), publisher: IOS Press

\bibitem{stromland_preregistration_2019}
Strømland, E.: Preregistration and reproducibility. Journal of Economic Psychology  \textbf{75},  102143 (2019)

\bibitem{tassa2018deepmind}
Tassa, Y., Doron, Y., Muldal, A., Erez, T., Li, Y., Casas, D.d.L., Budden, D., Abdolmaleki, A., Merel, J., Lefrancq, A., et~al.: Deepmind control suite. arXiv preprint arXiv:1801.00690  (2018)

\bibitem{tatman_practical_2018}
Tatman, R., VanderPlas, J., Dane, S.: A {Practical} {Taxonomy} of {Reproducibility} for {Machine} {Learning} {Research}. CEUR Workshop Proceedings  (2018)

\bibitem{trosten2023questionable}
Trosten, D.J.: Questionable practices in methodological deep learning research. In: Proceedings of the Northern Lights Deep Learning Workshop. vol.~4 (2023)

\bibitem{walonoski_synthea_2018}
Walonoski, J., Kramer, M., Nichols, J., Quina, A., Moesel, C., Hall, D., Duffett, C., Dube, K., Gallagher, T., McLachlan, S.: Synthea: {An} approach, method, and software mechanism for generating synthetic patients and the synthetic electronic health care record. Journal of the American Medical Informatics Association: JAMIA  \textbf{25}(3),  230--238 (2018)

\bibitem{wiggins_replication_2019}
Wiggins, B.J., Christopherson, C.D.: The replication crisis in psychology: {An} overview for theoretical and philosophical psychology. Journal of Theoretical and Philosophical Psychology  \textbf{39},  202--217 (2019)

\bibitem{wolff_probast_2019}
Wolff, R.F., Moons, K.G., Riley, R.D., Whiting, P.F., Westwood, M., Collins, G.S., Reitsma, J.B., Kleijnen, J., Mallett, S.: {PROBAST}: {A} {Tool} to {Assess} the {Risk} of {Bias} and {Applicability} of {Prediction} {Model} {Studies}. Annals of Internal Medicine  \textbf{170}(1),  51--58 (2019), publisher: American College of Physicians

\bibitem{xu_privacy-preserving_2021}
Xu, R., Baracaldo, N., Joshi, J.: Privacy-preserving machine learning: Methods, challenges and directions. arXiv preprint arXiv:2108.04417  (2021)

\bibitem{yildiz_reproducedpapersorg_2021}
Yildiz, B., Hung, H., Krijthe, J., Liem, C., Loog, M., Migut, G., Oliehoek, F., Panichella, A., Pawełczak, P., Picek, S., de~Weerdt, M., van Gemert, J.: {ReproducedPapers}.org: {Openly} {Teaching} and {Structuring} {Machine} {Learning} {Reproducibility}. Lecture Notes in Computer Science (including subseries Lecture Notes in Artificial Intelligence and Lecture Notes in Bioinformatics) pp. 3--11 (2021)

\bibitem{young_minatar_2019}
Young, K., Tian, T.: Minatar: An atari-inspired testbed for thorough and reproducible reinforcement learning experiments. arXiv preprint arXiv:1903.03176  (2019)

\end{thebibliography}

\end{document}